\documentclass[lettersize,journal,nosections]{IEEEtran}
\usepackage{amsmath,amsfonts}
\usepackage[linesnumbered,ruled,vlined]{algorithm2e}
\usepackage{array}
\usepackage[caption=false,font=normalsize,labelfont=sf,textfont=sf]{subfig}
\usepackage{textcomp}
\usepackage{stfloats}
\usepackage{url}
\usepackage{verbatim}
\usepackage{graphicx}
\usepackage{cite}
\usepackage{enumitem}
\usepackage{booktabs}
\usepackage{cleveref}
\crefformat{table}{#2Table~\Roman{#1}#3}

\usepackage{threeparttable}
\usepackage{multirow}

\usepackage{fancyhdr} 
\usepackage[switch]{lineno}
\usepackage{xcolor} 

\usepackage[switch]{lineno}  
\usepackage{xcolor}

\usepackage{cite}
\usepackage{amsmath,amssymb,amsfonts}
\usepackage{threeparttable}
\usepackage{multirow}
\usepackage{booktabs} 
\usepackage{algorithmic}
\usepackage{graphicx}
\usepackage{textcomp}
\usepackage{svg}
\usepackage{tcolorbox}
\usepackage{colortbl}   
\usepackage{booktabs}  
\usepackage{listings}       
\usepackage{xcolor}         
\usepackage[utf8]{inputenc} 
\usepackage{textcomp}       
\usepackage{beramono}       
\usepackage{caption} 

\definecolor{keywordpurple}{RGB}{170,0,120} 
\definecolor{codegreen}{rgb}{0.8588, 0.9922, 0.9255}   
\definecolor{codestring}{rgb}{0.5,0,0.5}    
\definecolor{codegray}{rgb}{0.5,0.5,0.5}    
\newcommand{\lstbg}[3][0pt]{{\fboxsep#1\colorbox{#2}{\strut #3}}}

\lstdefinestyle{VerilogNormalStyle}{
    language=Verilog,                   
    basicstyle=\ttfamily\scriptsize,  
    linewidth=\columnwidth,
    xleftmargin=1em,
    framexleftmargin=1em,
    xrightmargin=0pt,     
    keywordstyle=\color{keywordpurple}\bfseries, 
    commentstyle=\color{codegreen},     
    stringstyle=\color{codestring},     
    identifierstyle=\color{black},      
    numbers=left,                       
    numberstyle=\tiny\ttfamily\color{codegray}, 
    numbersep=5pt,                      
    captionpos=b, 
    showspaces=false,                   
    showstringspaces=false,             
    showtabs=false,                     
    frame=tb,                           
    framesep=3pt,                       
    rulesepcolor=\color{black},         
    tabsize=4,                          
    breaklines=true,                    
    postbreak=\mbox{\quad},
    breakatwhitespace=true,             
    inputencoding=utf8,                 
    literate=
    {_}{\_}{1}
    {~}{{\texttildelow}}1
    {<=}{{$<=$}}1
    {>=}{{$\ge$}}1
    {==}{{$=$}}1
    {!=}{{$\neq$}}1
    {&&}{{$\land$}}1
    {||}{{$\lor$}}1
    {|}{{\textbar}}1
    {&&&}{{$\land\land\land$}}1
    {|||}{{$\lor\lor\lor$}}1
    {<<}{{$\ll$}}1
    {>>}{{$\gg$}}1
    {->}{{$\rightarrow$}}1
    {|}{{\textbar}}1,
    morecomment=[f][\color{red}]{---}, 
    morecomment=[f][\color{codegreen}]{+++},
    morecomment=[f][\lstbg{red!20}]{-\ },
    morecomment=[f][\lstbg{codegreen}]{+\ },
    morecomment=[f][\color{blue}]{@@},
}

\definecolor{mygreen}{RGB}{192,255,192} 
\definecolor{myred}{RGB}{255,204,204}   
\definecolor{lightgray}{RGB}{230,230,230} 

\begin{document}

\title{Structural Mutation Based Differential Testing for FPGA Logic Synthesis Compilers}

\author{Zhihao Xu, Shikai Guo, Guilin Zhao, Siwen Wang, Qian Ma, Hui Li, Furui Zhan


\thanks{Z. Xu, S. Guo, G. Zhao, S. Wang, Q Ma, H Li, and F. Zhan are with the School of Information Science and Technology, Dalian Maritime University, Dalian, China and the Key Laboratory of Artificial Intelligence of Dalian, Dalian, China. E-mail: cemery.xzh@gmail.com, shikai.guo@dlmu.edu.cn, yelen3876@dlmu.edu.cn, wsw@dlmu.edu.cn, maqian@dlmu.edu.cn, li\_hui@dlmu.edu.cn, izfree@dlmu.edu.cn}

}
\newcommand{\rev}[1]{{\color{black} #1}}
\newcommand{\Zhihao}[1]{\color{red} Zhihao:#1}
\newcommand{\Component}{TPG }
\newcommand{\Componentt}{BDS }
\newcommand{\Componenttt}{ETC }
\newcommand{\hdlcoder}{HDL Coder }

\maketitle

\begin{abstract}
\rev{Field Programmable Gate Arrays (FPGAs) play a crucial role in Electronic Design Automation (EDA) applications, which have been widely used in safety-critical environments, including aerospace, chip manufacturing, and medical devices. 
A critical stage in FPGA development is logic synthesis, which enables developers to translate their software designs into hardware net lists. Logic synthesis compilers can facilitate the physical implementation of the chip and ensure the software designs efficiently mapped onto standard FPGA primitives.
However, bugs in FPGA logic synthesis compilers may lead to unexpected behaviors in target hardware net lists, which may cause potential security risks.
Therefore, it is critical to detect and repair such bugs in FPGA logic synthesis compilers. 
Existing methods mainly focus on automatic fuzz testing for FPGA logic synthesis compilers. But they are limited by the redundancy and uniformity of generated test case which caused the low efficiency in FPGA logic synthesis compiler testing.
To address this challenge, we propose a mutation-based method integrate with the Bayesian optimization to detect bugs in FPGA logic synthesis compilers. The method called SmootHDL, consists with three key components, the Test-program Generation Component (TPG), the Bayesian Diversity Selection Component (BDS), and the Equivalent Test Check Component (ETC). SmootHDL imitates from a set of test cases which are called original test cases.
To diversify the logic of original test cases, the \Component use a control-flow mutation strategy to generate new test cases which is  functional equivalent with the original test cases. Furthermore, for removing the redundancy of generated test cases, the \Componentt combine the Bayesian optimizer to construct a selector based on the control-flow diversity. Finally, the \Componenttt compares the original test cases and generated test cases. Theoretically, functional equivalent test case should have same output. If there are any disagree, SmootHDL will report and reproduce it.
We evaluated our method on the mainstream FPGA logic synthesis compilers. Experiments show that SmootHDL can generate more diverse test cases and detect more bugs compared with the latest testing methods for FPGA logic synthesis compilers. Over three months, SmootHDL found 16 bugs, 12 of which were confirmed by official technical support, and 6 have already been fixed.}
 
\end{abstract}

\begin{IEEEkeywords}
FPGA, Logic Synthesis Compiler, Equivalent Check
\end{IEEEkeywords}

\section{Introduction}

\begin{figure}[!t]
\centering
 \includegraphics[width=\linewidth]{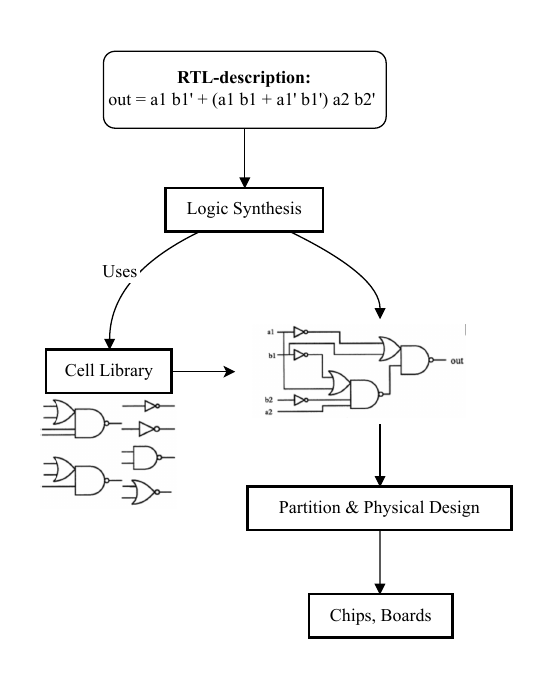}
\caption{The Conversion from HDL Code to Netlist}
\label{fig:1}
\end{figure}

\rev{Logic Synthesis is a crucial stage in the development of Field Programmable Gate Arrays (FPGAs). The FPGA logic synthesis compilers play critical role as a conduit from conceptual designs to executable digital circuits~\cite{HDL1,yosys,HDL2}.
These compilers allow engineers to convert high-level functionality descriptions, articulated in Hardware Design Language (HDL) codes like VHDL or Verilog, into gate-level net lists suitable for silicon chip fabrication. As shown as \figurename~\ref{fig:1}, 
this conversion is essential for creating complex Integrated Circuits (ICs). Within the context of Intel's FPGA Industrial Solutions Playbook, FPGA logic synthesis compilers ensure the ICs can be updated or reconfigured which avoid extensive hardware modifications~\cite{TCE1,TCE2,TCE3}. }

\rev{However, bugs in FPGA logic synthesis compilers may produce potential security risk~\cite{}.
Thus it is important to keep the accuracy and safety of FPGA logic synthesis compilers.
Recently, several methods have been proposed to detect bugs FPGA logic synthesis compilers. These methods mainly focused on using a fuzz test method to generate HDL test case for FPGA logic synthesis compilers.
Although they can detect numerous bugs in FPGA logic synthesis compiler, there also remain some limitations. 
The optimal method for testing FPGA logic synthesis compiler is EvoHDL~\cite{LegoHDL}, which generates Simulink Models guided by a AST and uses \hdlcoder~\cite{hdlcoder} to convert the Simulink Model to HDL test case for FPGA logic synthesis compiler testing. 
However, due to the limitation of Simulink Model library and the inherent optimization of \hdlcoder, the diversity of generated test cases from EvoHDL was limited. 
Another testing method is Verismith~\cite{verismith}, which uses an Abstract Syntax Tree (AST) generating method to create test cases and identify bugs. 
Despite its success in finding many bugs in FPGA logic synthesis compilers, Verismith's effectiveness is still limited by its simple corpus.

Thus, there remains two challenges in FPGA logic synthesis compiler testing.

\textbf{Challenge 1. Test Efficiency and Redundancy} Existing fuzzing-based methods~\cite{LegoHDL,verismith}, while effective in generating a large number of test cases, inherently suffer from a high degree of test case redundancy. The random nature of these approaches often leads to the repeated generation of semantically or structurally similar HDL code. This phenomenon, which we term stagnant test space exploration, results in a significant waste of computational resources and time. Instead of discovering new, valuable test cases that could uncover deep-seated compiler bugs, the testing process becomes trapped in local minima, repeatedly covering already-tested code paths.

\textbf{Challenge 2. Limited Test Case Diversity} Another major limitation of current methods is their reliance on simple corpus generation. Although this approach can produce syntactically valid code, the resulting test cases often exhibit semantic homogeneity. For example, these methods might generate countless test case of simple arithmetic logic, but they fail to create complex, bug-revealing structures like intricate state machines, deep pipelines, or nested loops. This lack of semantic diversity severely restricts the test's ability to expose bugs that only manifest under complex, real-world design conditions.

To address these challenge, we proposed SmootHDL, a mutation based FPGA logic synthesis compilers testing method. Our key insight is that control-flow mutation combined with Bayesian optimization expands the test space and reduces redundancy, leading to more diverse and efficient compiler testing.}
 
\rev{Specifically, SmootHDL consists of three main components: the Test-program Generation Component (TPG), the Bayesian Diversity Selection Component (BDS), and the Equivalent Test Check Component (ETC).
SmootHDL initiates with a set of test case which composed by Simulink Models, we call them seed Simulink Model. And for the HDL test case generated by the seed Simulink Model we call it original test case.
To generate diverse test cases in control flow, the \Component first collects all the Stateflow chart blocks in the seed Simulink Model. Then, the \Component extracts the state graph of the seed Simulink Model. Based on this state graph, the \Component performs structural mutations such as state duplication, path duplication, and transition expansion. Theoretically, these structural mutations should not alter the functionality of the seed Simulink Model but expand its state space. We call the new Simulink Model which has been gone through the structural mutations as mutant Simulink Model.
Then the \Componentt transform the mutant Simulink Model into HDL test case by \hdlcoder. But as we mentioned above, the inherent optimizations of \hdlcoder may cause different models to converge to similar HDL structures after translation. Hence by defining timing logic program distances, the \Componentt use Bayesian acquisition function to build a probability matrix for further test case selection. Since structural metamorphosis produces numerous functionally equivalent but structurally different variants, whose impact on timing complexity and diversity is uncertain. Bayesian inference naturally models this uncertainty and updates beliefs from observations, enabling efficient selection of valuable variants.
Finally, through the \Componenttt, if SmootHDL detects inconsistencies in the output of the variant HDL code with the original HDL code, it may indicate the presence of bugs in the logic synthesis tools.}

We evaluated our method on mainstream FPGA logic synthesis compilers. Experiments demonstrate that, within the same testing period, SmootHDL can find more bugs than existing methods. In three months of testing, we reported 16 unique bugs, 12 of which were confirmed by official developers, and 6 have already been fixed. Furthermore, SmootHDL has shown that it can generate more diverse test cases for FPGA logic synthesis compilers compared with existing methods. Finally, we verified the effectiveness of the \Component and \Componentt through ablation experiments.

\rev{In summary, the main contributions of this work are as follows:
\begin{itemize}
\item We introduce SmootHDL, a mutation-based method for FPGA logic synthesis compiler testing. To the best of our knowledge, this is the first approach that combines control-flow mutation with a Bayesian selector for FPGA logic synthesis compiler testing.
\item Extensive experiments are conducted to evaluate the bug-finding capability of SmootHDL. SmootHDL has detected 16 valid bugs, of which 12 bugs have been confirmed by official technical support and 6 have been fixed.
\item We release SmootHDL as a replication package for FPGA logic synthesis tool testing to foster reproducibility and facilitate further studies~\cite{HDLSmith}.
\end{itemize}
}
\begin{figure}[!t]
\centering
 \includegraphics[width=0.97\linewidth]{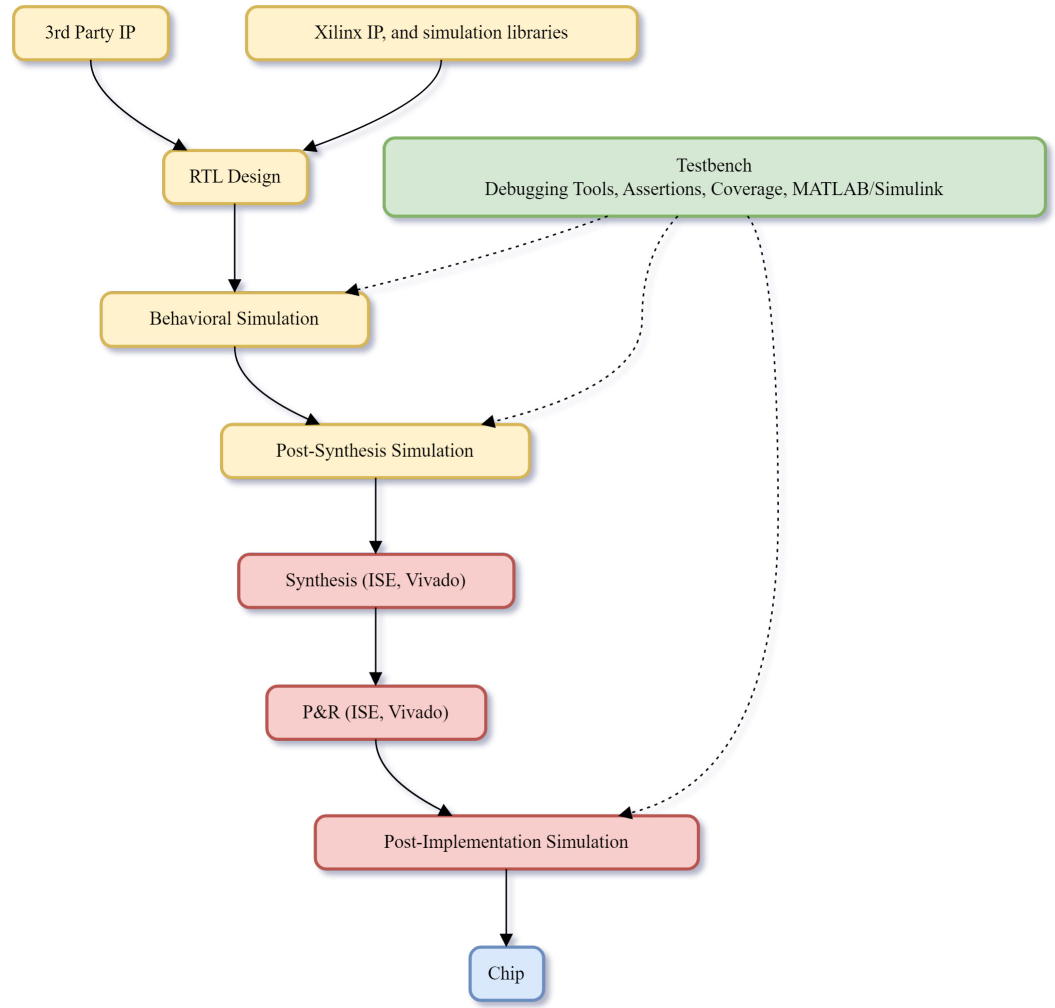}
\caption{The Process of Logic Synthesis}
\label{fig:2}
\end{figure}

\section{Background and Motivation}


\subsection{Preliminaries}

Logic synthesis tools play a crucial role in FPGA development by converting high-level design descriptions into gate-level representations that are optimized for FPGA hardware. 
These tools allow designers to concentrate on the algorithmic and architectural elements of their projects, handling the transformation of abstract specifications into logic circuits that can be implemented on hardware. 
They enhance the design by optimizing for speed, area, and power consumption, thus shortening the time-to-market and boosting the efficiency of FPGA-based applications. 
Additionally, these tools facilitate iterative design processes by enabling developers to modify their designs based on feedback from simulations and real-world tests, thereby ensuring that the final product achieves the best possible performance and resource efficiency.

The process as shown as~\figurename~\ref{fig:2}, begins with the input of a design in a Hardware Description Language (HDL), such as Verilog or VHDL. 
These HDL scripts describe the circuit's functionality but do not specify the physical implementation details. 
FPGA logic synthesis tools take these descriptions and perform optimizations and transformations, converting them into a mapped netlist of generic logic elements, such as lookup tables (LUTs) and flip-flops, which directly correspond to the resources available on the target FPGA.

Following the initial generation of the netlist, further optimizations are carried out to meet the specific requirements of the design. 
This includes optimizing the design for speed, power consumption, and area utilization. 
The FPGA logic synthesis tool applies various algorithms to rearrange, combine, or simplify logic to enhance performance and efficiency. 
Timing analysis is also performed to ensure that the circuit meets the desired operational frequency, and adjustments are made to the placement of logic elements and routing paths to minimize delays and avoid timing violations.

Finally, the logic synthesis process integrates the optimized netlist with the physical layout on the FPGA. 
This stage, known as place and route, involves placing the logic elements onto the physical grid of the FPGA and routing the interconnections between them. 
The output of this stage is a configuration file, often in the form of a bitstream, which can be loaded onto the FPGA to configure its logic blocks and interconnects to replicate the functionality defined by the HDL code. 
This file is crucial as it represents the final implementation of the design that will operate on the FPGA hardware.

\subsection{Illustrative Examples}

We present a real confirmed bug that existing methods failed to detect. In this example, we demonstrate the effectiveness of our structural mutation strategy. The bug involves a Vivado miscompilation issue, illustrated in List~\ref{lst:strategy1}, which shows the synthesized modules that trigger the error. Since the synthesized model typically contains thousands of modules, we highlight only the structural root cause.

\begin{lstlisting}[style=VerilogNormalStyle, caption= Example of Metamorphosis strategies in Logic Operation Optimization , label={lst:strategy1}]
localparam IDLE=2'b00, RUN=2'b01, RUN_DUP=2'b10;
always @* begin
    next_state = state;
    case (state)
        RUN: begin
            next_state = RUN;               
+            if (g_inhib)                   
+                next_state = RUN_DUP;       
+            if ((~rst_n) && reset_seen)     
+                next_state = RUN_DUP;       
        end
+        RUN_DUP: next_state = RUN;
    endcase
end
\end{lstlisting}

\textbf{State Duplication with Inhibited Branch:} SmootHDL duplicates a reachable state in the seed Simulink Model and inserts a complementary guard $g_{\mathit{inhib}}$ to control the cloned transition. Under default conditions $g_{\mathit{inhib}}=0$, the execution path remains identical to the original Simulink Model, ensuring functional equivalence. The cloned state introduces an unused structural branch in the control-flow graph, which increases the diversity and complexity of the generated HDL.

\textbf{Subsystem Wrapping and HDL Conversion:} During HDL conversion, the duplicated state and its inhibited transition are encapsulated into a new subsystem by HDL Coder. Each subsystem is then converted into an independent HDL file and linked into the main design. Although the inhibited branch should not affect functional behavior, the compiler is still required to process it during synthesis.

Unexpectedly, the synthesis tool crashed during optimization. Logically, the duplicated branch should have been eliminated, as its guard is never satisfied. However, the optimizer incorrectly retained this path, which involved the reset signal of the clock cycle. This mistake caused the synthesis engine to explore an invalid logic path. Under extreme timing conditions, the inhibited branch was incorrectly taken, leading to a loop in the control logic and ultimately triggering a crash.

This case demonstrates that structural mutations, such as state duplication with inhibited transitions, can expose deep bugs in FPGA logic synthesis tools that would not be discovered through simpler random fuzzing approaches.


\section{Methodology} 

\rev{In this section, we first present an overview of the SmootHDL framework and then explain its three major components: the Test Program Generation Component (\Component) the Bayesian Diversity Selection Component (\Componentt) and the Equivalent Test Check Component (\Componenttt).}

\subsection{Overview}

\begin{figure*}[htbp]
\centering
 \includegraphics[width=\linewidth]{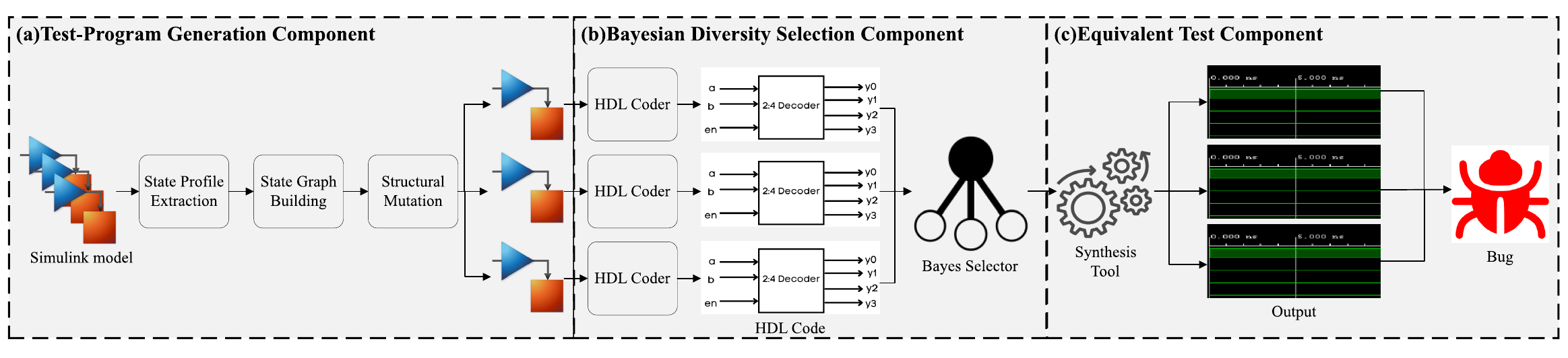}
\caption{The Framework of SmootHDL}
\label{fig:flow}
\end{figure*}

\begin{algorithm}[!t]
  \caption{SmootHDL}
  \label{alg:1}
  \KwIn{HDL Coder $L$, Seed Simulink Model $C$, Simulink parameters $P$, FPGA logic Synthesis tool $S$, HDL Code $H$, Mutant number MAX-ITER}
  \KwOut{Reported Bug}

  \SetKwFunction{FMain}{TEST}
  \SetKwFunction{FTPC}{TPG}
  \SetKwFunction{FBDS}{BDS}
  \SetKwProg{Fn}{Function}{:}{\KwRet}
  
  \Fn{\FMain{$S$, $C$, $P$, $L$, $H$}}{
    $H \gets L.\text{Compile}(C).\text{Execute}(P)$;
    
    $O \gets S.\text{Compile}(H)$;
    
    \For{$i=1$ \KwTo MAX-ITER}{
    
      $C' \gets \FTPC{C, P, L}$;
      
      $H'' \gets \FBDS{C', H}$;
      
      $O' \gets S.\text{Compile}(H'')$;
      
      \If{$O' \ne O$}{
      
        \Return \text{ReportBug}(S, $H''$);
        
      }
    }
  }
  
  \Fn{\FTPC{$C$, $P$, $L$}}{
    $V \gets \text{Profile}(C, P)$\;
    \tcp{Extract state graph}
    $A \gets \text{Extgraph}(C, P, V)$\;
    \tcp {structural mutations}\    
    $C' \gets \text{strumut}(C, A)$\;

    \KwRet $C'$;
  }
  
  \Fn{\FBDS{$C'$, $H$}}{
     
    $H' \gets L.\text{Compile}(C').\text{Execute}(P)$\;
    $B \gets \text{BayesSelector}(H', H)$\;    
    \If{$B \ne \text{FALSE}$}{
    
      $H'' \gets \text{Optimize}(H')$;
      
    }
    \KwRet $H''$;
  }
\end{algorithm}

\rev{An overview of the TRAGIC framework is illustrated in \figurename~\ref{fig:flow}. It takes Simulink Model as input and yields HDL test cases for bug detection.
To diversify the Simulink Models and further diverse the control-flow of generated HDL test cases, \Component uses a control-flow mutation for the input CPS. Specifically, \Component first collects and constructs the state graph in the input CPS. Then \Component use structural mutation strategies to mutate the input Simulink Model  based on the state graph. These structural mutation strategies will not change the original states of input Simulink Model but make the state flow more complex. Hence \Component makes the control flow more complex of HDL test cases generated by the input CPS.
After structural mutation, new Simulink Model which mutate from the input Simulink Model will be sent to the \Componentt to translate to HDL test cases. Recognize that the \hdlcoder may lead to homogenization for the generated test case. \Componentt uses a Bayesian selector to select the HDL test case based on the difference between the generated test case. It ensures that all selected HDL test cases are different in control flow structure. Finally, the \Componenttt validate output of the HDL test case generated by the input Simulink Model and generated HDL test case via simulation.  Theoretically, they should be consistent for the mutation and selection should not change the original function of the input Simulink Model. If there are any inconsistent SmootHDL report and reproduce it. We present the detailed algorithm in Algorithm~\ref{alg:1}, with lines 10-14 for the \Component module and lines 15-20 for the \Componentt module.
Next, we elaborate on the designs for the three modules of SmootHDL.}

\subsection{Test-Program Generation Component}

\rev{The \Component accepts a seed Simulink Model as input and produces structurally mutated variants that enlarge the state space while preserving functional behavior. The procedure consists of two stages: first, constructing the state graph of the input Simulink Model; second, applying structural mutation strategies on this graph to generate new Simulink Model  variants.

\subsubsection*{State Graph Construction}

In the first stage, the \Component detects all Stateflow chart blocks in the seed Simulink Model and extracts their states and transitions. The state graph is formally represented as
\begin{equation}
G = (S, T),
\end{equation}
where $S=\{s_1,s_2,\ldots,s_n\}$ is the set of states and $T=\{t_1,t_2,\ldots,t_m\}$ is the set of transitions. Each transition $t_k \in T$ is expressed as
\begin{equation}
t_k = (s_i, s_j, g),
\end{equation}
with $s_i$ as the source state, $s_j$ as the destination state, and $g$ as the transition guard. The graph $G$ captures the topology of the finite state machine and provides the basis for structural mutation.

\subsubsection*{Structural Mutation Strategies}

Based on the constructed state graph, the \Component applies three structural mutation strategies.

\textbf{State Duplication.} For a state $s \in S$, a cloned state $s'$ is created with identical actions such that
\begin{equation}
\psi(s') = \psi(s).
\end{equation}
The outgoing guard $g$ is decomposed into two complementary guards $g \land \neg g_{\mathit{inhib}}$ and $g \land g_{\mathit{inhib}}$, leading to $s$ and $s'$ respectively. When $g_{\mathit{inhib}}=0$, the execution remains equivalent to the original FSM, while the state set expands to $S' = S \cup \{s'\}$.

\textbf{Path Duplication.} For a path from $s_i$ to $s_j$ denoted as $P(s_i,s_j)$, a parallel path $P'(s_i,s_j')$ is created where $s_j'$ is a clone of $s_j$ satisfying $\psi(s_j')=\psi(s_j)$. Only one of the two paths is active during execution, preserving functional equivalence while enlarging the transition set.

\textbf{Transition Expansion.} For a transition $t=(s_i,s_j,g)$, an intermediate state $s_x$ is inserted to form
\begin{equation}
s_i \rightarrow s_x \rightarrow s_j,
\end{equation}
with $\psi(s_x)=\emptyset$. This guarantees that the composed transition preserves the semantics of $t$ while increasing the depth of the transition structure.

Through these mutations, the transformed Simulink Model maintains functional equivalence with the original while enlarging the state space. If the original model is represented as
\begin{equation}
M = (S, T),
\end{equation}
and the mutated model as
\begin{equation}
M' = (S', T'),
\end{equation}
then the relation
\begin{equation}
M' \equiv M \quad \text{and} \quad |S'| > |S|
\end{equation}
holds under default parameter settings. The enlarged state graph forces the logic synthesis compiler to process more complex control structures and exercise optimization strategies that remain untested with the seed Simulink Model.

\begin{figure}
    \centering
    \includegraphics[width=\linewidth]{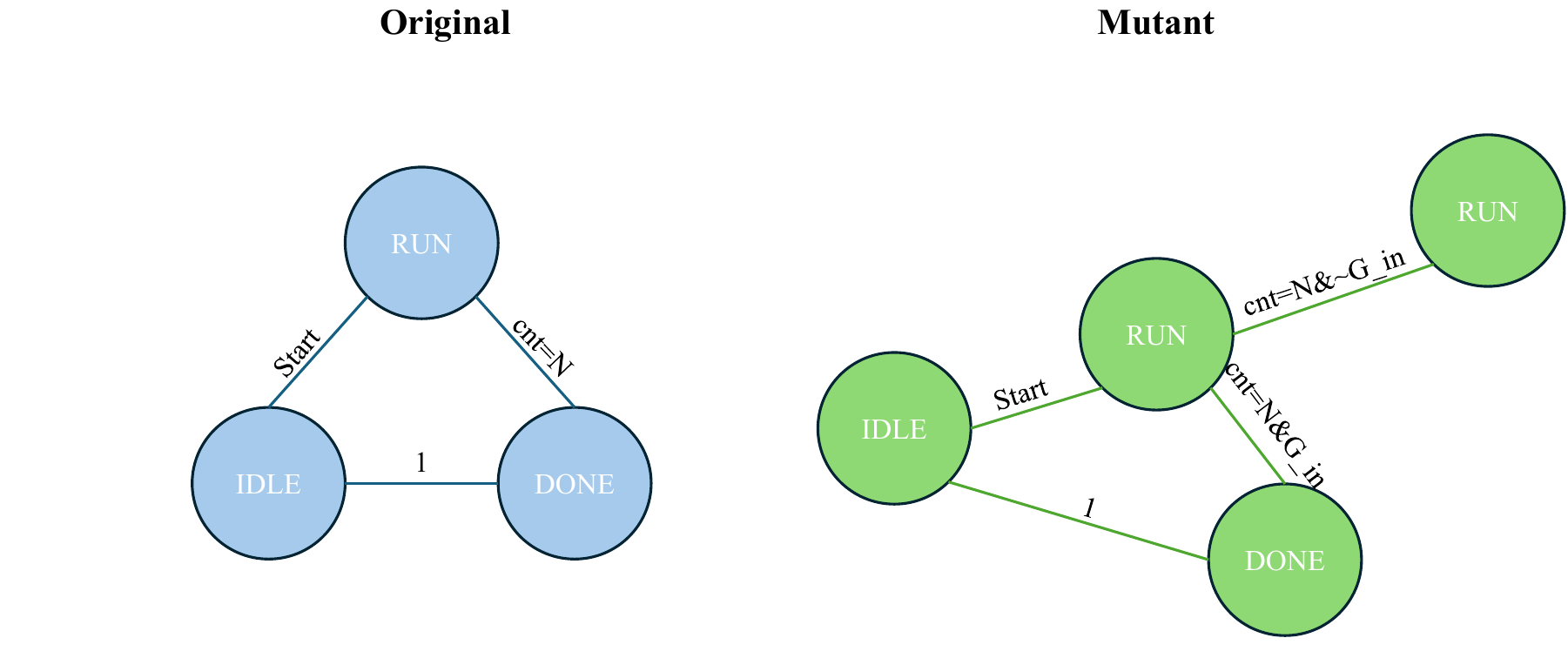}
    \caption{Example of State Duplication}
    \label{fig:example}
\end{figure}

Consider a simple Simulink Model of a counter controller with three states $IDLE$, $RUN$, and $DONE$. The original transitions are $IDLE \xrightarrow{start} RUN$, $RUN \xrightarrow{cnt==N} DONE$, and $DONE \xrightarrow{1} IDLE$, as shown in \figurename~\ref{fig:example} (left). 
After applying state duplication, a new state $DONE'$ is introduced with $\psi(DONE')=\psi(DONE)$. The transition from $RUN$ is split into two complementary guards, $cnt==N \land \neg g_{\mathit{inhib}}$ leading to $DONE$ and $cnt==N \land g_{\mathit{inhib}}$ leading to $DONE'$. The resulting FSM has four states while preserving functional equivalence, as illustrated in Figure~\ref{fig:example} (right). The functional equivalence is guaranteed by two conditions. First, the guards $cnt==N \land \neg g_{\mathit{inhib}}$ and $cnt==N \land g_{\mathit{inhib}}$ are mutually exclusive, ensuring that only one of the two transitions can be taken at runtime. Second, the cloned state $DONE'$ is defined to have identical semantics as $DONE$, that is $\psi(DONE')=\psi(DONE)$. As a result, for any input sequence, the execution trace of the mutated FSM can be mapped to an equivalent trace in the original FSM, thereby preserving determinism and observable behavior while introducing additional structural complexity.}

\subsection{Bayesian Diversity Selection Component}

\rev{The \Componentt first convert the Simulink Model variant generated by the Test-Program Generation component into HDL Code through HDL Coder.Secondly, we recognize that the inherent optimizations of HDL Coder may cause different Simulink Models to generate structurally similar HDL programs, which reduces the effectiveness of testing. To address this issue, the Bayesian Diversity Selection (BDS) component selects program variants with higher structural diversity by combining program distance with a Bayesian selector.} 

\begin{algorithm}[!t]

\caption{Bayesian Diversity Selection}
\label{alg:bayesian_diversity_selection}
\KwIn{Seed Simulink Model $c$, HDL Coder $L$, List of program variants $V$, Prior probabilities}
\KwOut{Selected variants for testing}

\SetKwFunction{FBDS}{FBDS}
\SetKwFunction{FDist}{Dist}
\SetKwFunction{FProb}{Prob}
\SetKwProg{Fn}{Function}{:}{\KwRet}

\Fn{\FBDS{$c$, $L$, $V$, $P(V)$}}{
    $H \gets L.\text{Compile}(c)$\;
    
    $d(V_i) \gets \text{FDist}(H, V_i)$\;
    
    \ForEach{$V_i \in V$}{
        $d(V_i) \gets \text{FDist}(H, V_i)$\;
        
        $P(V_i|D) \gets \text{FProb}(V_i, d(V_i), P(V))$\;
    }
    
    Selected $\gets$ Sort and select top variants based on $P(V_i|D)$\;
    
    \KwRet \text{SelectedVariants}\;
}

\Fn{\FDist{$H$, $V_i$}}{
    \KwRet $\sqrt{(v_{1H} - v_{1V_i})^2 + (c_{1H} - c_{1V_i})^2 + (s_{1H} - s_{1V_i})^2}$\;
}

\Fn{\FProb{$V_i$, $d$, $P(V)$}}{
    $\text{numerator} \gets P(D|V_i) \cdot P(V_i)$\;
    
    $\text{denominator} \gets \sum_{j=1}^{|V|} P(D|V_j) \cdot P(V_j)$\;
    
    \KwRet $\frac{\text{numerator}}{\text{denominator}}$\;
}

\end{algorithm}










The Bayesian selector is a decision-making tool based on Bayesian decision theory and is used to make optimal choices under uncertainty. It uses prior knowledge and newly acquired data to update probability estimates to guide decision-making. The basic strategy of Bayesian selectors is to select variants with high temporal complexity and program complexity. SmootHDL first define the program distance $d$ between two programs as follow.

\begin{equation}
d(P_1, P_2) = \sqrt{(v_1 - v_2)^2 + (c_1 - c_2)^2 + (s_1 - s_2)^2}
\label{eq:program_distance}
\end{equation}

Formula~\eqref{eq:program_distance}  defines the Euclidean distance \(d(P_1, P_2)\) between two programs \(P_1\) and \(P_2\) based on three metrics: the number of net variables \(v\), the number of connections \(c\), and the number of process structures \(s\). Each metric contributes to the squared difference in their respective counts between the two programs, summed and then square-rooted to produce the distance. This measure quantifies the structural similarity between two programs, where a smaller distance indicates greater similarity in their composition and complexity.

Then SmootHDL utilize Bayesian inference. Prior probability \( P(V_i) \) for each variant \( V_i \) represents our initial belief about its complexity. The likelihood \( P(D|V_i) \), indicates the probability of observing specific complexity metrics \( D \) given a variant \( V_i \). The posterior probability \( P(V_i|D) \), which updates our belief in the complexity of each variant based on observed data, is computed using formula \eqref{eq:bayes1}:

\begin{equation}
P(V_i|D) = \frac{P(D|V_i) \cdot P(V_i)}{\sum_{j=1}^{N} P(D|V_j) \cdot P(V_j)}
\label{eq:bayes1}
\end{equation}

Specifically, given a variant \( V_i \), the choice to accept the variant \( V_i \) depends on two aspects: the program distance \( P(D|V_i) \) and timing complexity ratio \( P(V_i) \)  of the program from the original program, and the program distance \( P(D|V_j) \) and timing complexity ratio \( P(V_i) \) of the program to other programs.

Timing complexity in digital circuits refers to the measurement of the time requirements for signal transmission and processing, with a focus on the maximum time taken for a signal to travel from one logic unit to another along the longest path, known as the critical path. This maximum delay, constrained by the circuit’s clock frequency and synchronization operations, directly impacts the circuit's performance and operational frequency. The timing complexity is essential in ensuring that all signal operations within the circuit are completed within one clock cycle to avoid timing errors and performance bottlenecks.
The calculation formula of timing complexity is as shown in the formula~\eqref{eq:timingcomplexity}

\begin{equation}
P(V_i) = \max \left( \sum_{i=1}^{n} t_{i} \right)
\label{eq:timingcomplexity}
\end{equation}
where \( P(V_i) \) represents the maximum delay across the circuit, essentially the delay along the critical path; \( n \) is the number of components on that path; \( t_i \) denotes the delay contributed by the \(i\)-th component. This calculation ensures that all components complete their operations before the arrival of the next clock signal, thereby maintaining circuit integrity and efficiency.

In this way, SmootHDL calculate the acceptance probability \( P(V_i|D) \) of all variant \( V_i \)  and form these probabilities into a one-dimensional matrix. This simplifies the selection of variants. If the test program setting requires multiple variants, SmootHDL select those with the highest acceptance probability from the matrix. Through Bayesian selectors, SmootHDL efficiently generates more complex HDL code for variant test programs.

\subsection{Equivalent Check Component}

\rev{Since the netlist files generated by logic synthesis are usually not directly comparable, SmootHDL has chosen simulation  to verify the consistency of the netlist file. In addition, Yosys provides the \texttt{abc} engine for equivalence checking of netlist files, which we use as an auxiliary method to support verification, although it is not the primary approach in our framework.

The \Componenttt consists of two sub-processes using different simulation tools and comparing original and variant model outputs. The output of the simulation tool is a waveform containing high and low levels(as shown in the \figurename~\ref{fig:flow}), so SmootHDL print the output as a 0/1 level signal format and compare them bit by bit. 

For different simulation tools, SmootHDL use the same file to generate netlist files through different FPGA logic synthesis compilers, and use the same stimulus file to simulate them to compare their outputs. Formula~\eqref{eq:SimExec} represents the execution of simulation \( \mathcal{S}_j \) on netlist \( \mathcal{N} \) using stimuli \( \sigma \), resulting in output \( \mathcal{O}_j \). Each simulation tool might use a slightly different engine or algorithm, and \( \mathcal{O}_j \) captures the output of the \( j \)-th tool.

\begin{equation}
\mathcal{S}_j(\mathcal{N}, \sigma) = \mathcal{O}_j \quad \text{for} \, j = 1, 2, \ldots, k
\label{eq:SimExec}
\end{equation}

After simulations, outputs are compared to assess the consistency of the netlist's behavior across different simulation platforms based on Formula~\eqref{eq:ConsistencyCheck}. This formula ensures that all outputs \( \mathcal{O}_i \) are identical for every pair of simulations, confirming that the netlist behaves consistently irrespective of the simulation tool used. If the output is inconsistent, SmootHDL will report the issue. We will manually check the issues reported by SmootHDL and report to developers.

\begin{equation}
\text{Consistent}(\mathcal{O}_1, \mathcal{O}_2, \ldots, \mathcal{O}_k) = \bigwedge_{i=1}^{k-1} \bigwedge_{j=i+1}^{k} (\mathcal{O}_i = \mathcal{O}_j)
\label{eq:ConsistencyCheck}
\end{equation}


For comparing original and variant model outputs, SmootHDL prints the signal as a 0/1 electrical signal, and compares the simulation result output after the original Simulink Model is converted to HDL code and synthesized bit by bit with the simulation result output after the variant Simulink Model is converted into HDL code and synthesized. Theoretically the results should not change since the inserted module will not be executed and translated. But if the results are inconsistent, SmootHDL will report the Bugs. We will manually check these issues to find the root causes and submit them to the corresponding software developers.}


\section{Evaluation}

In this section, four experiments are conducted to evaluate the effectiveness of SmootHDL. Specifically, our evaluation aims at answering the following Research Questions (RQs).

\textbf{RQ1:} How is the bug-finding capability of SmootHDL compared to the state-of-the-art methods?

\textbf{RQ2:} Can SmootHDL detect new FPGA logic synthesis compiler bugs?

\textbf{RQ3:} How diverse is the HDL code generated from SmootHDL compared to the state-of-the-art methods?

\textbf{RQ4:} How effective is the \Componentt of SmootHDL?

In our experiments, RQ1 and RQ2 are used to evaluate the bug-finding capability of SmootHDL compared to the state-of-the-art approaches. RQ3 and RQ4 are employed to evaluate the code generation strategy of the code generation component and the diverse variant generation component.

\subsection{State of The Art Method}

There are various testing methods available for FPGA logic synthesis compilers, such as VerilogHammer, Verismith, and LegoHDL. However, since VerilogHammer does not support the latest Verilog standard, we have selected Verismith and LegoHDL as our baseline methods.

\subsection{Evaluation Setup}

SmootHDL has been developed using MATLAB and Python, and both our code and experimental data are publicly accessible on GitHub~\cite{HDLSmith}. The evaluation of SmootHDL was conducted on a computer running the Ubuntu 22.04 operating system, equipped with an Intel Core i9 CPU @ 2.10GHz, and 128GB of memory. Based on the recommendations of existing work, we focus on the size of generated HDL files being in the 700-100 line range~\cite{verismith}.

In order to clearly illustrate the root cause of bugs, we utilize automated reduction approaches to simplify the HDL code that triggers the bug. 
This process enables developers to quickly comprehend and rectify bugs. 
Specifically, we employ a technique resembling the binary search method. 
Leveraging the AST extracted from the HDL code, we iteratively delete portions of the code until the erroneous use case can no longer be minimized.
To avoid reporting duplicate bugs, we manually use failed assertions and back-trace to detect duplicates. 
When two bugs have same failed assertion or back-trace, we consider them as duplicates. 
Finally, we report the detected bugs which are verified as non-duplicates as new issues to developers. 

\subsection{Answer to RQ1}

\textbf{Approach.} To evaluate the effectiveness of SmootHDL, we compare the bug-finding capability of SmootHDL with the state-of-the-art methods Verismith~\cite{verismith} and LegoHDL~\cite{LegoHDL}, since finding more bugs within a time period is the main objective of these approaches. In the experiment, we detect bugs on both the recently released version of FPGA logic synthesis compiler include Vivado, yosys, Iverilog and Quartus. We set a single testing period of two weeks for FPGA logic synthesis compiler, that is, every method tests these FPGA synthesis compilers for two weeks.

\begin{table}[!ht]
    \centering
    \caption{Bugs detected by Our method, LegoHDL, and Verismith}
    \begin{tabular}{p{2cm}ccc}
    \toprule
        \textbf{} & \textbf{Known} & \textbf{New} & \textbf{Total} \\
        \midrule
        LegoHDL & 2 & 1 & 3 \\
        Verismith & 0 & 1 & 1 \\
        \textbf{Our Method} & \textbf{2} & \textbf{4} & \textbf{6} \\ 
    \bottomrule
    \end{tabular}
    \label{tab:exp1}
\end{table}

\textbf{Results.} As illustrated in Table \ref{tab:exp1}, bugs detected in our experiment are categorized into new bugs (\emph{New}) and previously identified ones (\emph{Known}), with the latter being duplicates from the bug repository. Notably, Table \ref{tab:exp1} demonstrates that SmootHDL significantly outperforms Verismith and LegoHDL in bug detection capabilities. Over four weeks, SmootHDL identified six bugs in these FPGA logic synthesis compilers, comprising two new bugs and four known ones. In contrast, LegoHDL detected three bugs—two known and one new—while Verismith identified only one known bug. This occurs because SmootHDL is capable of generating HDL test cases that are more complex than those produced by the baseline method. During the translation of the Simulink Model to HDL code by the HDL coder, each subsystem is converted into a separate HDL file with defined reference relationships. SmootHDL enhances the complexity of these relationships by integrating complex code branches into the Simulink Model, thereby challenging the FPGA logic synthesis compiler to employ more sophisticated optimization strategies for the HDL files. Additionally, the Bayesian selector intensively tests the compiler by selecting increasingly complex HDL code scenarios for processing. All these lead to SmootHDL's bug discovery efficiency being better than the baseline method.

\begin{center}
    \begin{tcolorbox}[colback=gray!10,
                      colframe=black,
                      width=\linewidth,
                      arc=1mm, auto outer arc,
                      boxrule=0.5pt,
                     ]
    \textbf{Conclusion.} SmootHDL shows better bug-finding capability than LegoHDL and Verismith for testing FPGA logic synthesis compiler. In two week testing period SmootHDL found 6 bugs, which 4 is unknown before. 
    \end{tcolorbox}

\end{center}


\subsection{Answer to RQ2}

\begin{figure}[!t]
\centering
 \includegraphics[width=0.95\linewidth]{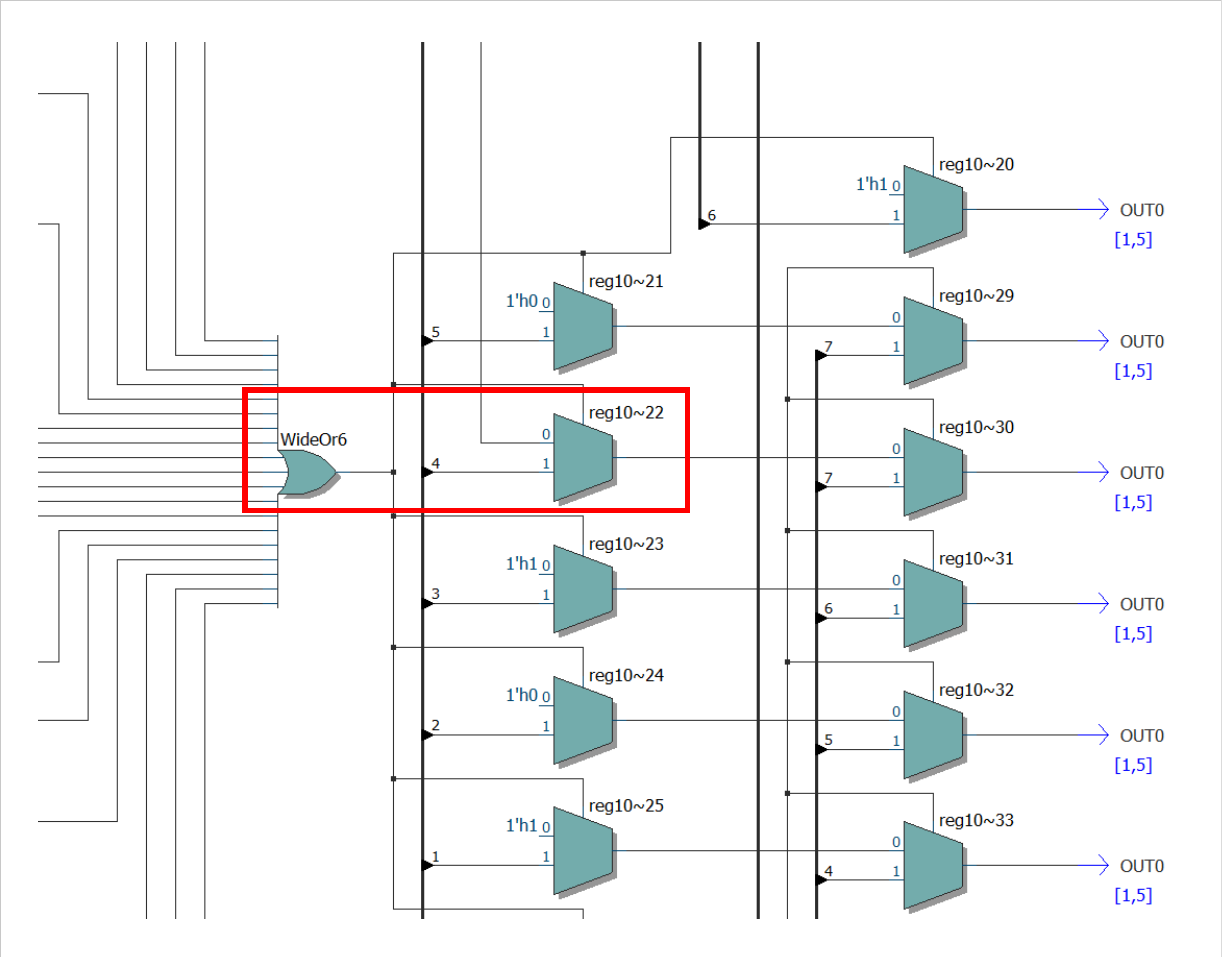}
\caption{FPGA Logic Synthesis Compiler Bug Found via Our Method}
\label{fig:bug2}
\end{figure}

\textbf{Approach.} Over a two-month period from May to July 2024, we extensively tested SmootHDL to assess its bug-finding capabilities. Unfortunately, we could not validate the latest version of Quartus due to licensing issues. However, we employed the latest versions of other FPGA logic synthesis compilers for testing. Each identified issue with a logic synthesis compiler was reported to the corresponding software community, and we engaged with developers there to gather feedback.

\textbf{Results.} As depicted in Table \ref{tab:bug}, our approach successfully identified 16 bugs within a three-month timeframe. The official websites of various logic synthesis tools, including Iverilog, Vivado, Yosys, and Quartus, have classified each issue as either Confirmed as New, Confirmed as Known, or Unconfirmed in Doubt (Pending Verification). New indicates bugs previously unrecognized by developers. Known refers to issues already identified as bugs by developers. Pending represents issues that developers believe can be circumvented through specific standardized procedures. To facilitate replication of our findings, we have made the HDL code files that triggered these bugs available on GitHub \cite{HDLSmith}.
Notably, four new bugs were detected that had not been identified by previous methods. \figurename \ref{fig:bug2} illustrates one such case revealed by SmootHDL. In this example, our method duplicated a feasible state and inserted an inhibited branch, which after Simulink-to-HDL translation was encapsulated into a new subsystem. The resulting HDL introduced a redundant WideOr6 logic component connected to multiple multiplexers. Although the inhibited guard should prevent this path from being activated, Vivado incorrectly retained it during optimization. As a result, the synthesis tool attempted to propagate signals through this unused logic cone, which involved reset-related nets, eventually leading to a failure in the optimization stage.
Conventional methods such as Verismith and LegoHDL are unable to expose this Bugs because they rarely generate designs with such redundant and inhibited branches at the control-flow level. In contrast, SmootHDL’s structural mutations and Bayesian-guided selection deliberately increase both control-flow and timing complexity, making it more effective at revealing hidden synthesis bugs of this kind.

\begin{table*}[!t]
  \centering
  \caption{Bugs found by Our Method}
  \begin{threeparttable}
    \begin{tabular}{lllll}
    \toprule
    ID    & Summary & Status & Type  & Software \\
    \midrule
    Yosys-1 & support error for Aggregate Initialization and Replication& Verified & C     & Yosys \\
    Yosys-2 & Buffer Overflow Triggered by Inadequate Input Validation & Verified & C     & Yosys \\
    Yosys-3 & Incorrect State Machine Transition Due to Race Condition & Verified & C     & Yosys \\
    Yosys-4 & Memory Leak Arising from Unreleased Dynamic Allocations & Pending & M     & Yosys \\
    Iverilog-1 & SystemVerilog reference error & Verified & M     & Iverilog \\
    Vivado-1 & Synthesis crash under unsigned delay of specific symbol & Verified & C     & Vivado \\
    Vivado-2 & Synthesis failed and crash(HARTGLADDGen:regenerate(bool)) & Verified & C     & Vivado \\
    Vivado-3 & Vivado synthesis failed (ConstPropagate) & Verified & C     & Vivado \\
    Vivado-4 & Vivado synthesis failed (NANname) & Verified & C     & Vivado \\
    Vivado-5 & Synthesis failed caused by ConstProp::reconnect(NNet,NNet). & Verified & C     & Vivado \\
    Vivado-6 & Synthesis failed caused by inspection function. & Verified & C     & Vivado \\
    Vivado-7 & Reference relationship error caused synthesis failed & Pending & C     & Vivado \\
    Vivado-8 & Synthesis failed caused by NNet function & Verified & C     & Vivado \\
    Vivado-9 & Data Corruption from Improper Handling of Concurrency & Verified & M     & Vivado \\
    Vivado-10 & Inconsistent Data Integrity Caused by Faulty Synchronization & Pending & M     & Vivado \\
    Quartus-1 & Global RAM value reset inconsistencies & Pending& C     & Quartus \\
    \bottomrule
    \end{tabular}%
     \begin{tablenotes}
        \footnotesize
        \item[1] There are two types of status feedback from the official bug report (i.e., $Verified$ = newly confirmed bug, $pending$ = Pending verification). There are two types of bugs (i.e., Type) in our reported bugs: crash bugs ($C$) and miscompilation bugs ($M$).
    \end{tablenotes}
    \end{threeparttable}
\label{tab:bug}
\end{table*}%

\begin{center}
    \begin{tcolorbox}[colback=gray!10,
                      colframe=black,
                      width=\linewidth,
                      arc=1mm, auto outer arc,
                      boxrule=0.5pt,
                     ]
    \textbf{Conclusion.} SmootHDL has demonstrated effectiveness in detecting bugs in FPGA logic synthesis compilers. Within a two-month period, it identified 16 valid bugs, among which 4 were confirmed as new by the official developers. 
    \end{tcolorbox}

\end{center}

\subsection{Answer to RQ3}

\textbf{Approach.} We employ SmootHDL, LegoHDL, and Verismith to generate HDL code test cases, each consisting of 700–1000 lines and totaling 1000 test cases. We compare their complexity and generation time using three key indicators derived from the Abstract Syntax Tree (AST) and synthesis reports. The first metric, the count of blocks, tallies the blocks in the generated files, where a higher block count reflects greater structural complexity. The second metric, the count of lines or connections, indicates that a higher number of connections increases the likelihood of encountering synthesis issues. The third metric is the critical path delay obtained from synthesis timing analysis, where larger delay values represent deeper logic paths and more complex timing structures that can stress synthesis tools more thoroughly.

\begin{figure}[!t]
\centering
 \includegraphics[width=0.95\linewidth]{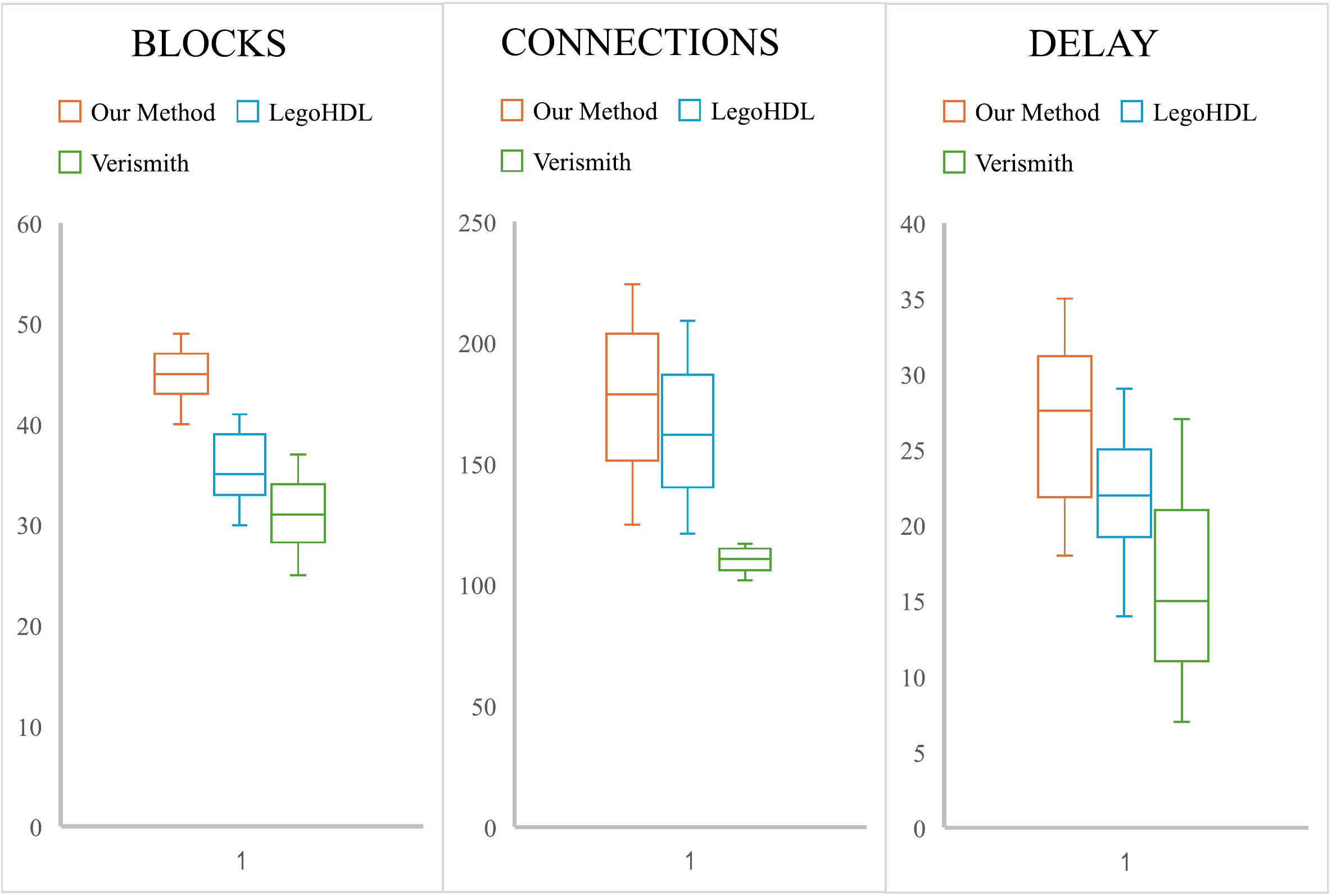}
\caption{Blocks, connections and references generated by Our Method, LegoHDL and Verismith}
\label{fig:exp3}
\end{figure}

\textbf{Results.} \textit{Numbers of Blocks and Connections.}
Blocks and connections are vital components of HDL code, crucial for depicting its characteristics. As shown in \figurename \ref{fig:exp3}, SmootHDL generates a range of [40,49] blocks and [125,226] connections, surpassing the ranges produced by Verismith \cite{verismith} ([25,37] blocks and [102,117] connections) and LegoHDL ([30,41] blocks and [120,210] connections). This increased complexity arises as SmootHDL continuously selects more intricate HDL codes for testing, based on a Bayesian selector, enhancing its bug detection efficiency over the baseline methods.
\textit{Delay.} \figurename \ref{fig:exp3} reports the distribution of critical path delay for LegoHDL, Verismith, and our method. The median delay of LegoHDL is 22.0, while Verismith is lower at 14.5. Our method shows a higher median of 27.6. In terms of range, LegoHDL spans 14–29, Verismith 7–26, and our method 18–35.
At the upper end, our method produces a larger proportion of high-delay test cases. Specifically, 66\% of cases exceed 25s and 32\% exceed 30s, compared with 14\% and none for LegoHDL, and 2\% and none for Verismith. At the lower end, 75\% test cases of Verismith and 27\% test cases of LegoHDL fall below 20s, while our method shows only 11\% in this range.
These observations indicate that, the control-flow mutations and Bayesian-guided selection shift the delay distribution upward, producing HDL test cases with deeper logic paths and therefore higher timing complexity than the two baselines.

\begin{center}
    \begin{tcolorbox}[colback=gray!10,
                      colframe=black,
                      width=\linewidth,
                      arc=1mm, auto outer arc,
                      boxrule=0.5pt,
                     ]
    \textbf{Conclusion.} Compared with Verismith and LegoHDL, SmootHDL generates HDL test cases with higher structural complexity and richer reference relationships. These results indicate that SmootHDL can provide greater diversity in generated test cases, which is valuable for exercising FPGA logic synthesis compilers.
    \end{tcolorbox}

\end{center}

\subsection{Answer to RQ4}

\begin{figure}[!t]
\centering
 \includegraphics[width= 0.90\linewidth]{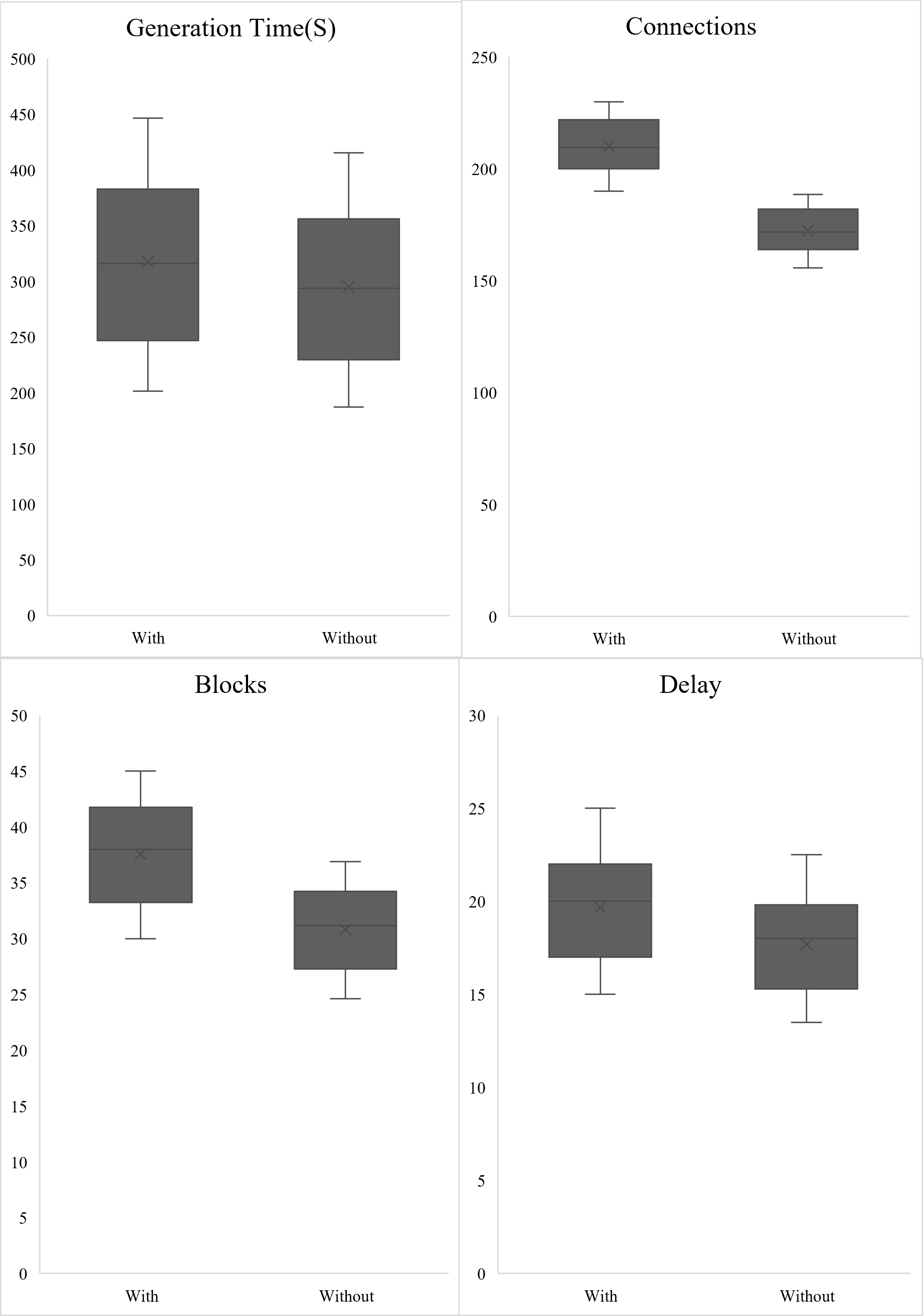}
\caption{Blocks, Connections and Delay Comparison with Bayesian Selector and without Bayesian Selector}
\label{fig:exp4}
\end{figure}

\textbf{Approach.} We conducted an ablation to quantify the effect of the Bayesian selector in SmootHDL. Two settings were evaluated: with selector and without selector. Each setting produced 100 HDL test cases within 700–1000 lines under the same tool chain and hardware configuration. We recorded generation time for each case. Structural metrics were extracted from the synthesized AST, which is block count and connection count. Timing complexity was measured as the critical-path delay reported by the synthesis timing analysis. All other parameters of the generator were fixed so that the selector was the only changing factor.

\textbf{Results.}  The selector shifts the distributions of structure and timing upward with a modest increase in generation time. From the box-plots in \figurename \ref{fig:exp4}, the median number of connections increases from about 175 without the selector to about 210 with the selector, an increase of roughly twenty percent; the inter quartile ranges show little overlap, indicating a consistent effect. The median number of blocks rises from about 31 to about 38, an increase of roughly twenty to twenty-five percent, again with clearly separated quartiles. The median critical-path delay increases from about 18 ns to about 20 ns, an increase of roughly ten to twelve percent; the distributions partly overlap, which suggests a moderate timing-complexity gain. Generation time shows a small upward shift, with the median moving from about 300 s to about 320–330 s, around seven to ten percent. Taken together, these observations indicate that the selector tends to guide the generator toward HDL designs with higher structural richness and somewhat deeper logic paths, while the additional runtime is limited under this experimental budget.

\begin{center}
    \begin{tcolorbox}[colback=gray!10,
                      colframe=black,
                      width=\linewidth,
                      arc=1mm, auto outer arc,
                      boxrule=0.5pt,
                     ]
    \textbf{Conclusion.} The Bayesian selector contributes to the effectiveness of SmootHDL by guiding the selection of variants with greater structural complexity and diversity, while maintaining comparable generation efficiency.
    \end{tcolorbox}

\end{center}

\section{Threats to Validity}

\textbf{Internal Threats.} The primary threat to the internal validity of our method lies in the accuracy of SmootHDL, as well as the reproduced models Verismith \cite{verismith} and LegoHDL \cite{LegoHDL}. To mitigate errors in SmootHDL, we meticulously double-checked our code. Additionally, we reproduced Verismith and LegoHDL by closely following the open-source code provided in their original publications, and we have thoroughly verified their accuracy. Furthermore, the model checking capabilities of Simulink HDL Coder can lead to increased processing times when generating large volumes of code. To address this, we optimized block placement to minimize model checking duration. Moving forward, we plan to intensify our testing of the FPGA logic Synthesis compiler by focusing on the generation of more complex test cases.

\textbf{External Threats.} A significant threat to the external validity of SmootHDL is the potential for triggering duplicate bugs during testing. To mitigate this, we have implemented strategies to reduce bug-triggering variants by excising blocks unrelated to the core issues, which enhances our ability to accurately understand and analyze bugs. Based on the developers's verification, the bugs we reported are currently not duplicates, suggesting that our approach is effective in minimizing this threat.
\rev{Furthermore, the effectiveness of the Bayesian selector depends on the richness of the candidate pool and the granularity of the diversity features. When the available transformations yield candidates that are semantically similar at the FSM level (for example, simple pipelines or largely stateless datapaths), the posterior can concentrate early and the selector may provide limited benefit. In SmootHDL we mitigate this risk by enriching the pool through control-flow and timing oriented mutations, including state or path cloning, inhibited transitions, and scheduling variations. And we plan to study stopping rules that detect pool homogeneity and trigger automatic pool refresh, so that the selector remains effective when diversity declines.}

\section{Related Work}
\label{sec:related-work}
\subsection{FPGA Logic Synthesis Testing}
In the field of FPGA logic synthesis compiler testing, Verismith~\cite{verismith} and LegoHDL emerge as the primary methodologies for generating HDL code and testing logic synthesis tools. Verismith~\cite{verismith} functions as a Verilog program generator that crafts random behavioral Verilog code devoid of undefined values, based on predefined parameter configurations. Nevertheless, Verismith's~\cite{verismith} capability to generate complex HDL code is limited as it exclusively produces Verilog, a single hardware design language.

To overcome these limitations and enhance the thoroughness of FPGA logic synthesis compiler testing, Xu et al. introduced LegoHDL~\cite{LegoHDL}, which aims to ensure the accuracy and correctness of FPGA logic synthesis processes. LegoHDL employs \hdlcoder to generate HDL code by constructing a Simulink Model. It extracts the AST from the HDL code and orchestrates the construction of the Simulink Model to facilitate interaction between Simulink Model construction and HDL code generation. Finally, LegoHDL utilizes Metamorphic testing to synthesize the HDL code using various synthesizers and verify the equivalence of the generated netlist files. However, LegoHDL still faces challenges with insufficient diversity, and its singular Simulink Model   construction approach does not adequately address deeper bugs.

Another significant tool in this domain is VlogHammer~\cite{VlogHammer}, a Verilog fuzzer specifically designed for testing major commercial FPGA logic synthesis compilers and various simulators. As of now, VlogHammer~\cite{VlogHammer} has identified approximately 75 defects. However, in contrast to Verismith~\cite{verismith}, VlogHammer does not generate multi-model programs and lacks support for key behavioral-level Verilog constructs, such as \texttt{always} blocks. Similarly, VlogHammer~\cite{VlogHammer} also limits its output exclusively to Verilog.

Additionally, random Verilog generators such as VERGEN~\cite{VEGEN} exist, producing behavioral-level Verilog by randomly assembling high-level logic blocks, including state machines, MUXes, and shift registers. These generators, however, depend on predefined definitions to construct their structures, leading to a limited diversity of generated configurations. This limitation reduces their ability to thoroughly test a wide spectrum of Verilog structure combinations. 


In recent years, Large Language Models (LLMs) like the open-source CodeGen-16B, used in VeriGen~\cite{VeriGen} for test case generation, have shown promising results in FPGA logic synthesis compiler testing, boasting a 41\% improvement in generating syntactically correct Verilog code over traditional pre-trained LLM methods. Despite this, VeriGen’s focus is not primarily on testing; its generation is limited to a dataset sourced from GitHub and textbooks, which may not thoroughly challenge FPGA compilers and lacks the capability to generate stimulus files, thereby restricting test accuracy. Other approaches, such as BetterV~\cite{BetterV}, are focused on optimizing Verilog code and reducing runtime. Currently constrained by the accuracy of the corpus and the scale of model generation, LLMs prove insufficient for effective HDL generation testing.

Furthermore, numerous efforts have been dedicated to enhancing the reliability of FPGA development compilers. For example, Yann Herklotz et al.\cite{vericert} have concentrated on enhancing the stability of high-level synthesis (HLS) tools through formal verification, introducing Vericert, a formally verified HLS tool. Zewei Du et al.\cite{zedu} have investigated the use of fuzz testing to comprehensively evaluate HLS tools by deploying a vast array of valid C programs. Despite these contributions, a gap remains in the quality and complexity of the generated files, predominantly focusing on a single programming language. Consequently, we have embarked on more comprehensive testing of FPGA logic synthesis compilers to affirm their stability and reliability.

\subsection{Metamorphic testing in Compiler testing}

Metamorphic testing generates variants of test cases considered equivalent, based on program inputs. It identifies bugs by comparing the outcomes of these variants \cite{ChenHHXZ0X16,R31,R32,R50}. Metamorphic testing has been extensively verified with widely used compilation tools, including GCC, and this method has identified over 1,000 bugs \cite{orion,athena,Hemers,R19}. In compiler testing, three EMI-based Metamorphic testing mutation approaches are employed: Orion \cite{orion}, Athena \cite{athena}, and Hermes \cite{Hemers}. Orion specifically targets dynamically dead program regions, randomly pruning unexecuted statements to generate variant programs \cite{orion}. Athena inserts or deletes code in dead code regions under varying inputs \cite{athena}. Unlike Orion and Athena, which are limited to mutating dead code regions, Hermes \cite{Hemers} can alter both live and dead code regions to generate equivalent variants.

Metamorphic testing has also been applied in other fields. For example, SLEMI~\cite{SLEMI} proposed by Shaful et al. uses strategies such as deleting dead code to generate Simulink Model  variants to thoroughly test the Simulink compiler. Guo et al. proposed COMBAT~\cite{combat}, which uses the idea of live code mutation to generate Simulink Model  variants in order to more thoroughly test the Simulink compiler. In addition, the idea of Metamorphic testing has also been applied at the project level. Wu et al. proposed Mantra~\cite{Mantra} to apply mutation testing to HDL and achieved good results.
 

\section{Conclusions and Future Work}

In this paper, we proposed a novel method for FPGA logic synthesis compiler testing, named SmootHDL. This method incorporates a Test-Program Generation component and a Bayesian selection component, enabling the generation of diverse HDL code for testing the FPGA logic synthesis compiler. Over a three-month period, our method successfully identified 16 bugs, 12 of which have been confirmed by developers. Going forward, we plan to explore how to integrate large models to guide the generation of even more diverse test cases, further enhancing our testing of FPGA logic synthesis compilers.


\section*{Acknowledgment}
This work was supported by the National Natural Science Foundation of China (No.62472062), the Dalian Excellent Young Project (2022RY35), the Fundamental Research Funds for the Central Universities (No.3132024257).



%

\bibliographystyle{unsrt}
\bibliography{mytse}

\begin{thebibliography}{10}

\bibitem{HDL1}
Michiel Ligthart, Karl Fant, Ross Smith, Alexander Taubin, and Alex Kondratyev.
\newblock Asynchronous design using commercial hdl synthesis tools.
\newblock In {\em Proceedings Sixth International Symposium on Advanced Research in Asynchronous Circuits and Systems (ASYNC 2000)(Cat. No. PR00586)}, pages 114--125. IEEE, 2000.

\bibitem{yosys}
Clifford Wolf, Johann Glaser, and Johannes Kepler.
\newblock Yosys-a free verilog synthesis suite.
\newblock In {\em Proceedings of the 21st Austrian Workshop on Microelectronics (Austrochip)}, page~97, 2013.

\bibitem{HDL2}
R~Dekker, M~Ligthart, and L~Lapides.
\newblock Hdl synthesis for fpga design.
\newblock {\em Electronic Engineering}, 66(814), 1994.

\bibitem{TCE1}
Brunel~Happi Tietche, Olivier Romain, and Bruce Denby.
\newblock Sparse channelizer for fpga-based simultaneous multichannel drm30 receiver.
\newblock {\em IEEE Transactions on Consumer Electronics}, 61(2):151--159, 2015.

\bibitem{TCE2}
Matías~J. Garrido, Fernando Pescador, M.~Chavarrías, P.~J. Lobo, and César Sanz.
\newblock A high performance fpga-based architecture for the future video coding adaptive multiple core transform.
\newblock {\em IEEE Transactions on Consumer Electronics}, 64(1):53--60, 2018.

\bibitem{TCE3}
Xue Liu, Ze-Ke Wang, and Qing-Xu Deng.
\newblock Fpga implementation of a reconfigurable channelization for simultaneous multichannel drm30/fm receiver.
\newblock {\em IEEE Transactions on Consumer Electronics}, 63(1):1--9, 2017.

\bibitem{LegoHDL}
Zhihao Xu, Shikai Guo, Guilin Zhao, Peiyu Zou, Xiaochen Li, and He~Jiang.
\newblock A novel hdl code generator for effectively testing fpga logic synthesis compilers, 2024.

\bibitem{hdlcoder}
Hdl coder.
\newblock \url{https://www.mathworks.com/products/hdl-coder.html}, 2023.

\bibitem{verismith}
Yann Herklotz and John Wickerson.
\newblock Finding and understanding bugs in {FPGA} synthesis tools.
\newblock In Stephen Neuendorffer and Lesley Shannon, editors, {\em {FPGA} '20: The 2020 {ACM/SIGDA} International Symposium on Field-Programmable Gate Arrays, Seaside, CA, USA, February 23-25, 2020}, pages 277--287. {ACM}, 2020.

\bibitem{HDLSmith}
Lsc-fuzz.
\newblock \url{https://github.com/cemery123/FPGA_Logic_Synthesis_Compilers_Testing}, 2024.

\bibitem{VlogHammer}
Vloghammer.
\newblock \url{https://github.com/YosysHQ/VlogHammer}, 2019.

\bibitem{VEGEN}
Boris Ratchev, Mike Hutton, Gregg Baeckler, and Babette van Antwerpen.
\newblock Verifying the correctness of fpga logic synthesis algorithms.
\newblock In {\em Proceedings of the 2003 ACM/SIGDA Eleventh International Symposium on Field Programmable Gate Arrays}, FPGA '03, page 127–135, New York, NY, USA, 2003. Association for Computing Machinery.

\bibitem{VeriGen}
Shailja Thakur, Baleegh Ahmad, Hammond Pearce, Benjamin Tan, Brendan Dolan-Gavitt, Ramesh Karri, and Siddharth Garg.
\newblock Verigen: A large language model for verilog code generation.
\newblock {\em ACM Transactions on Design Automation of Electronic Systems}, 29(3):1--31, 2024.

\bibitem{BetterV}
Zehua Pei, Hui{-}Ling Zhen, Mingxuan Yuan, Yu~Huang, and Bei Yu.
\newblock Betterv: Controlled verilog generation with discriminative guidance.
\newblock {\em CoRR}, abs/2402.03375, 2024.

\bibitem{vericert}
Michalis Pardalos, Yann Herklotz, and John Wickerson.
\newblock Resource sharing for verified high-level synthesis.
\newblock In {\em 30th {IEEE} Annual International Symposium on Field-Programmable Custom Computing Machines, {FCCM} 2022, New York City, NY, USA, May 15-18, 2022}, pages 1--6. {IEEE}, 2022.

\bibitem{zedu}
Yann Herklotz, Zewei Du, Nadesh Ramanathan, and John Wickerson.
\newblock An empirical study of the reliability of high-level synthesis tools.
\newblock In {\em 29th {IEEE} Annual International Symposium on Field-Programmable Custom Computing Machines, {FCCM} 2021, Orlando, FL, USA, May 9-12, 2021}, pages 219--223. {IEEE}, 2021.

\bibitem{ChenHHXZ0X16}
Junjie Chen, Wenxiang Hu, Dan Hao, Yingfei Xiong, Hongyu Zhang, Lu~Zhang, and Bing Xie.
\newblock An empirical comparison of compiler testing techniques.
\newblock In Laura~K. Dillon, Willem Visser, and Laurie~A. Williams, editors, {\em Proceedings of the 38th International Conference on Software Engineering, {ICSE} 2016, Austin, TX, USA, May 14-22, 2016}, pages 180--190. {ACM}, 2016.

\bibitem{R31}
William~M. McKeeman.
\newblock Differential testing for software.
\newblock {\em Digital Technical Journa}, 10(1):100--107, 1998.

\bibitem{R32}
Christopher Lidbury, Andrei Lascu, Nathan Chong, and Alastari~F. Donaldson.
\newblock Many-core compiler fuzzing.
\newblock {\em ACM SIGPLAN Notices}, 50(6):65--76, 2015.

\bibitem{R50}
Yibiao Yang, Yuming Zhou, Hao Sun, Zhendong Su, Zhiqiang Zuo, Lei Xu, and Baowen Xu.
\newblock Hunting for bugs in code coverage tools via randomized differential testing.
\newblock {\em International Conference on Software Engineering (ICSE 2019)}, pages 488--499, 2019.

\bibitem{orion}
Vu~Le, Mehrdad Afshari, and Zhendong Su.
\newblock Compiler validation via equivalence modulo inputs.
\newblock In Michael F.~P. O'Boyle and Keshav Pingali, editors, {\em {ACM} {SIGPLAN} Conference on Programming Language Design and Implementation, {PLDI} '14, Edinburgh, United Kingdom - June 09 - 11, 2014}, pages 216--226. {ACM}, 2014.

\bibitem{athena}
Vu~Le, Chengnian Sun, and Zhendong Su.
\newblock Finding deep compiler bugs via guided stochastic program mutation.
\newblock In Jonathan Aldrich and Patrick Eugster, editors, {\em Proceedings of the 2015 {ACM} {SIGPLAN} International Conference on Object-Oriented Programming, Systems, Languages, and Applications, {OOPSLA} 2015, part of {SPLASH} 2015, Pittsburgh, PA, USA, October 25-30, 2015}, pages 386--399. {ACM}, 2015.

\bibitem{Hemers}
Chengnian Sun, Vu~Le, and Zhendong Su.
\newblock Finding compiler bugs via live code mutation.
\newblock In Eelco Visser and Yannis Smaragdakis, editors, {\em Proceedings of the 2016 {ACM} {SIGPLAN} International Conference on Object-Oriented Programming, Systems, Languages, and Applications, {OOPSLA} 2016, part of {SPLASH} 2016, Amsterdam, The Netherlands, October 30 - November 4, 2016}, pages 849--863. {ACM}, 2016.

\bibitem{R19}
Yixuan Tang, He~Jiang, Zhide Zhou, Xiaochen Li, Zhilei Ren, and Weiqiang Kong.
\newblock Detecting compiler warning defects via diversity-guided program mutation.
\newblock {\em IEEE Transactions on Software Engineering}, pages 1--1, 2021.

\bibitem{SLEMI}
Shafiul~Azam Chowdhury, Sohil~Lal Shrestha, Taylor~T. Johnson, and Christoph Csallner.
\newblock {SLEMI:} equivalence modulo input {(EMI)} based mutation of {CPS} models for finding compiler bugs in simulink.
\newblock In Gregg Rothermel and Doo{-}Hwan Bae, editors, {\em {ICSE} '20: 42nd International Conference on Software Engineering, Seoul, South Korea, 27 June - 19 July, 2020}, pages 335--346. {ACM}, 2020.

\bibitem{combat}
Shikai Guo, He~Jiang, Zhihao Xu, Xiaochen Li, Zhilei Ren, Zhide Zhou, and Rong Chen.
\newblock Detecting simulink compiler bugs via controllable zombie blocks mutation.
\newblock In Abhik Roychoudhury, Cristian Cadar, and Miryung Kim, editors, {\em Proceedings of the 30th {ACM} Joint European Software Engineering Conference and Symposium on the Foundations of Software Engineering, {ESEC/FSE} 2022, Singapore, Singapore, November 14-18, 2022}, pages 1061--1072. {ACM}, 2022.

\bibitem{Mantra}
Jiang Wu, Yan Lei, Zhuo Zhang, Xiankai Meng, Deheng Yang, Pan Li, Jiayu He, and Xiaoguang Mao.
\newblock Mantra: Mutation testing of hardware design code based on real bugs.
\newblock In {\em 2023 60th ACM/IEEE Design Automation Conference (DAC)}, pages 1--6, 2023.

\end{thebibliography}
	

%








\end{document}